\documentclass[12pt, a4paper]{iopart}

\usepackage[T1]{fontenc}
\usepackage[english]{babel}

\usepackage{graphicx}
\usepackage{xcolor}
\usepackage{multirow}
\usepackage{booktabs}
\usepackage{xurl}

\expandafter\let\csname equation*\endcsname\relax
\expandafter\let\csname endequation*\endcsname\relax
\usepackage{mathtools}
\usepackage{iopams,amssymb,bm}

\usepackage[numbers,square,sort&compress]{natbib}

\definecolor{xgreen}{HTML}{009E73}
\definecolor{xverm}{HTML}{D55E00}
\definecolor{xmag}{HTML}{CC79A7}
\definecolor{xlinkcolor}{HTML}{0072B2}

\usepackage[
  colorlinks,
  linkcolor=magenta,
  citecolor=xverm,
  filecolor=xgreen,
  urlcolor=cyan
]{hyperref}

\makeatletter
\@addtoreset{equation}{section}
\renewcommand\theequation{\thesection.\arabic{equation}}

\makeatother

\makeatletter
\long\def\@makecaption#1#2{%
  \vskip\abovecaptionskip
  {\small
   \setlength{\parindent}{0pt}%
   \setlength{\baselineskip}{11pt}%
   \noindent{\bf #1.} #2\par}
  \vskip\belowcaptionskip
}
\makeatother

\usepackage{etoolbox}
\AtBeginEnvironment{thebibliography}{%
  \small
  \setlength{\itemsep}{0pt}
  \setlength{\parskip}{0pt}
  \setlength{\baselineskip}{0.80\baselineskip}
}

\makeatletter
\renewcommand{\subsection}{\@startsection{subsection}{2}{\z@}
  {-3.25ex plus -1ex minus -.2ex}
  {1.5ex plus .2ex}
  {\normalfont\normalsize\bfseries}}
\makeatother

\usepackage[scaled=0.95]{inconsolata}
\newcommand{\IMRPhenomD}{\texttt{IMRPhenomD}}
\newcommand{\aLIGOZeroDetHighPower}{\texttt{aLIGOZeroDetHighPower}}

\begin{document}


\title[Template bank using low-discrepancy sequences]{Stochastic template banks for GW searches using low-discrepancy sequences}

\author{Tarun Kumar and Anand S. Sengupta}
\address{Dept. of Physics, Indian Institute of Technology Gandhinagar, Gujarat 382055, India.}
\ead{kumar.tarun@alumni.iitgn.ac.in, asengupta@iitgn.ac.in}

\begin{abstract}
Matched filtering remains the most sensitive method for detecting gravitational waves from compact binary coalescences. The efficiency of such searches depends on how well a discrete template bank covers the underlying parameter space. Conventional geometric, stochastic, and hybrid placement methods can lead to uneven coverage and redundant templates in higher dimensions. Hybrid methods are generally the most efficient among these, while stochastic methods are simpler to implement, particularly when the parameter-space metric is difficult to compute. In practice, both approaches rely on uniform random sampling, 
which often requires a large number of proposal points to achieve adequate coverage.
We find that stochastic template banks constructed using low-discrepancy sequences achieve comparable recovery fractions 
while requiring 27.5\% fewer proposal points in two dimensions, \textcolor{black}{about 12\% fewer in three dimensions and 17\% fewer in a four-dimensional eccentric equal-spin toy bank.} \textcolor{black}{The final template count is much less sensitive to the proposal sampling, consistent with the metric-volume constraints of the covering problem.}
%
\textcolor{black}{We introduce a slab-wise coarse prefilter to accelerate
candidate rejection for strongly anisotropic metrics. Applied to the 4D toy
bank, it gives a total bank-construction speed-up of about $1.28\times$.}
%
The primary benefit of low-discrepancy sampling is therefore a reduction in
the size of the initial proposal set, leading to lower memory usage and reduced
bookkeeping during bank generation. 
\textcolor{black}{These results
show that low-discrepancy sampling provides a simple improvement to stochastic
template-bank generation for current and future gravitational-wave searches.}
\end{abstract}

\section{INTRODUCTION} 
\label{sec:intro}



Gravitational waves (GWs) are propagating perturbations of spacetime generated by accelerating mass distributions, with compact binary coalescences among their most prominent astrophysical sources. Owing to the weak coupling of gravity to matter, the strains induced by astrophysical GWs at terrestrial detectors are typically extremely small, necessitating highly sensitive interferometric observatories, including LIGO~\citep{Aasi_2015}, Virgo~\citep{Acernese_2015}, and KAGRA~\citep{PhysRevD.88.043007}. 
The first direct detection of gravitational waves, GW150914~\citep{PhysRevD.93.122003}, observed by the two LIGO detectors, marked the beginning of gravitational-wave astronomy. The signal was attributed to the inspiral and merger of a spinning binary black hole system with inferred source-frame component masses of $36^{+5}_{-4}\, M_\odot$ and $29^{+4}_{-4}\, M_\odot$~\citep{PhysRevD.93.122003} respectively.


From the first observing run (O1) through the second part of the fourth
observing run (O4b), the cumulative GWTC-5.0 catalog~\citep{LVK:GWTC5} contains 390
compact-binary coalescence candidates with
$p_{\rm astro}\geq 0.5$, where $p_{\rm astro}$~\citep{PhysRevD.108.083043} is the inferred
probability that a candidate is of astrophysical origin
rather than terrestrial noise.

Looking ahead, the detection rate is expected to grow significantly with next-generation observatories such as LISA~\citep{Armano2019}, the Einstein Telescope~\citep{Abac2026}, and Cosmic Explorer~\citep{Evans:2021gyd}, whose enhanced low-frequency sensitivity will extend the observable horizon and enlarge the accessible source population. The addition of LIGO-India~\citep{Iyer2011LIGOIndia, ligoindia2023} will further strengthen the ground-based detector network by improving detection efficiency and source localization capabilities~\citep{PhysRevD.109.044051}, while proposed decihertz mission concepts such as IndIGO-D~\citep{Sharma:2026tff} could enable multiband observations and early-warning localization of compact binary coalescences.

%
The detection of gravitational waves from coalescing compact binaries~\citep{LIGOScientific:2016aoc} relies on matched filtering~\citep{PhysRevD.44.3819, PhysRevD.46.5236, PhysRevD.49.1707}, in which the detector data are correlated with a bank of modeled waveforms (templates) that discretely cover the intrinsic parameter space, with each template representing a specific point in that space.
%
The template bank is constructed such that, for any potential signal in the intrinsic parameter space, it contains at least one template within a distance ${\mathcal{D}_{\max} = \sqrt{1-\mathcal{M}_{\min}}}$ determined by a chosen minimal match $\mathcal{M}_{\min}$. The value of $\mathcal{M}_{\min}$ sets the tradeoff between detection sensitivity and computational cost: a lower value permits larger mismatches between the signal and the nearest template, reducing the recovered signal-to-noise ratio (SNR) and hence the detection efficiency, whereas a higher value improves signal recovery at the expense of a larger number of templates and increased computational cost for both offline analyses and low-latency search pipelines. In this work, we adopt the standard choice of $\mathcal{M}_{\min}=0.97$, which, under the usual scaling arguments, implies an expected loss of about 10\% in detection rate due to the discrete coverage of the template bank~\citep{PhysRevD.60.022002}.

Over the years, several approaches have been developed to make template bank generation as efficient as possible. Geometric methods based on lattice placement or metric tiling~\citep{PhysRevD.60.022002, PhysRevD.76.102004} achieve near-optimal coverage when the parameter space is flat and low-dimensional. However, the intrinsic parameter space is not globally flat, and curvature effects introduce coverage gaps that require significant fine-tuning. This makes geometric methods difficult to generalize to higher dimensions, while the absence of a reliable metric for complex waveform models further restricts their applicability.

Stochastic methods~\citep{Babak2008, PhysRevD.80.104014} provide a flexible alternative to geometric placement techniques. In \textcolor{black}{the conventional} bottom-up formulation, candidate templates are proposed sequentially, with the first proposal accepted immediately and each subsequent proposal added to the bank only if it lies at least $\mathcal{D}_{\max}$ away from all previously accepted templates. This process continues until the average acceptance ratio, computed over a fixed number of recent proposals, falls below a prescribed threshold, typically $10^{-4}$.

An alternative stochastic-placement method, introduced in~\citep{PhysRevD.95.104045} as a top-down approach, begins with an oversampled proposal set $\mathcal{U}$ containing $|\mathcal{U}|$ candidate points distributed across the parameter space.
Here $|\mathcal{U}|$ denotes the number of candidate points in the proposal set. 
The template bank is then constructed iteratively by selecting a point at random from $\mathcal{U}$, adding it to the bank $\mathcal{T}$, and removing all points lying within a distance $\mathcal{D}_{\max}$ of the selected point. This procedure continues until $\mathcal{U}$ is exhausted. While the final template counts in the two approaches are nearly identical, the top-down method was shown in~\citep{PhysRevD.95.104045} to be approximately twice as fast. 
In either approach, the distance criterion may be evaluated using the parameter-space metric, when available, or through direct computation of the match between waveforms.

Finally, hybrid banks~\citep{PhysRevD.95.104045, PhysRevD.99.024048} combine geometric placement with stochastic sampling to achieve more efficient template placement. In this geometric-random approach, one begins with an initial oversampled set of random proposals, $\mathcal{U}$, distributed across the parameter space. A randomly selected proposal is used as a seed, and only those neighboring points of an $A_n^*$ lattice that lie within the prescribed search space are added to the template bank $\mathcal{T}$, under the assumption that the parameter space is locally flat. Redundant proposals in $\mathcal{U}$ lying within a distance $\mathcal{D}_{\max}$ of these accepted templates are discarded. This lattice-based expansion is repeated by treating each newly added template as a seed until no further valid lattice points can be added. The remaining proposals in $\mathcal{U}$, corresponding to regions not efficiently covered by the local geometric approximation owing to curvature in the parameter space or boundary effects, are then incorporated through the stochastic component of the algorithm.

\textcolor{black}{Direct comparisons in Refs.~\citep{PhysRevD.95.104045, PhysRevD.99.024048} showed that hybrid geometric-random banks can achieve comparable fitting-factor coverage with fewer templates and substantially lower construction cost than purely stochastic banks. The present work instead holds the top-down stochastic placement algorithm fixed in order to isolate the effect of replacing random-uniform proposals by LDS proposals.}

In a naive implementation, the top-down stochastic algorithm can scale as
$\mathcal{O}(N^2)$ in the worst case, since candidate rejection may require
comparing selected templates with a large fraction of the proposal set.
\textcolor{black}{Alternative spatial-indexing approaches have also been developed to reduce this cost, including the neighboring-cell method of Fehrmann and Pletsch~\citep{FehrmannPletsch2014}, which uses spatial hashing to restrict distance calculations to nearby cells.}
Tree-based implementations, such as KD-trees~\citep{Bentley1975}, substantially reduce this cost and typically give scaling closer to $\mathcal{O}(N\log N)$ for well-behaved
point sets in moderate dimensions. Nevertheless, the size
$N=|\mathcal{U}|$ of the initial proposal set remains a practical bottleneck,
both in runtime and memory usage. This motivates the use of proposal
strategies that reduce the required oversampling without compromising
coverage.

The use of low-discrepancy sequences in this context was first explored by Manca \& Vallisneri~\citep{PhysRevD.81.024004} in the framework of self-avoiding random coverings, where improved covering efficiency relative to uniform random sampling was demonstrated in flat Euclidean spaces. Here, we extend that idea to the construction of stochastic template banks for the intrinsic parameter space of compact binary gravitational-wave signals, where the underlying signal manifold is curved rather than flat.

We show that low-discrepancy sequences significantly reduce $|\mathcal{U}|$ while achieving recovery fractions comparable to those obtained using larger proposal sets generated from uniform random sampling.
The use of LDS sampling also yields a modest reduction in the total number of templates, typically in the range $[-1.10,\ -0.42]\%$ in two dimensions and \textcolor{black}{$[-1.14,\ -0.83]\%$} in three dimensions. 
These results demonstrate that for the parameter spaces studied here, comparable recovery fractions are obtained with 10–30\% fewer proposal points.


Low-discrepancy sampling is not expected to significantly reduce the final template count for a fixed minimal match. 
The number of templates required to cover a parameter space is determined primarily by the metric volume of the signal manifold, the chosen minimal match, and the efficiency of the underlying covering.
Instead, the principal advantage of LDS lies in reducing the size $|\mathcal{U}|$ needed to achieve a given coverage. This reduction lowers memory requirements and improves the efficiency of nearest-neighbour searches used during stochastic bank construction, benefits that become increasingly important in higher-dimensional parameter spaces.

This paper is organized as follows. 
Sec.~\ref{sec:Low_discrepancy_seq} provides a brief introduction to Low Discrepancy Sequences (LDS), covering their generation and their advantages for efficient space-filling. Sec.~\ref{sec:Parameter_space_metric} reviews the metric on the signal manifold, establishing the notation used throughout the paper and introducing the key concepts. This sets the stage for Sec.~\ref{sec:Template bank construction}, where we present the stochastic 
top-down algorithm employed in this work, along with key terminology such as fitting factors used in the subsequent analysis. Finally, Sec.~\ref{sec:Results} compiles the results by generating template banks 
across the parameter sets defined in Table~\ref{tab:sets}, followed by a fitting factor study based on $10^5$ random injections, with performance 
assessed via cumulative distribution functions (CDFs) of fitting factors. \textcolor{black}{The 2D and 3D analyses use the \IMRPhenomD{} waveform model~\citep{PhysRevD.93.044007}, while the 4D eccentric equal-spin bank uses \texttt{TaylorF2Ecc}~\citep{Moore:2016suz, lalsuite}. For the \IMRPhenomD{} banks, the upper frequency cutoff is set to the innermost stable circular orbit frequency $f_{\mathrm{isco}}$ corresponding to the total mass of the compact binary system.}

\section{LOW-DISCREPANCY SEQUENCES}
\label{sec:Low_discrepancy_seq}
In this section, we provide a brief overview of low-discrepancy sequences (LDS). 
These are deterministic sequences designed to sample a domain more uniformly than pseudo-random points. 
The first $N$ elements of such a sequence define a proposal set with lower discrepancy than a typical pseudo-random sample of the same size, making them useful for generating well-distributed candidate points in stochastic template-bank construction.

These sequences have been extensively investigated and shown to be beneficial for Monte Carlo integration in higher dimensions compared to random uniform sampling.\citep{Brandimarte2014LDS, Niederreiter1992, DickPillichshammer2010}.
The two low-discrepancy sequences used in this paper are the \textit{Halton}~\citep{Halton1960} and
\textit{Sobol} sequences \citep{SOBOL196786}. Both can be viewed as extensions of the one-dimensional
\textit{van der Corput} sequence \citep{VanderCorput1935}. 
The construction begins by expressing a non-negative integer $n$ in base $b$ (usually a prime number) as $n = (\ldots d_4 d_3 d_2 d_1 d_0)_b$, or equivalently,
\begin{equation}
n = \sum_{k=0}^{m} d_k\, b^{k}.\end{equation}
The digits are then reversed with respect to the radix point to map the number into the unit interval, yielding
$
h = (0.d_0 d_1 d_2 d_3 \ldots )_b.
$
Accordingly, the $n$-th element of the van der Corput sequence in base $b$ is given by
\begin{equation}
V(n,b) = \sum_{k=0}^{m} d_k\, b^{-(k+1)}.
\label{eq:Van_der-corput}
\end{equation}


A Halton sequence generalizes the one-dimensional van der Corput construction to higher dimensions by assigning each coordinate its own prime base: the first coordinate uses base $2$, the second base $3$, the third base $5$, and so on \citep{Halton1960}. \textcolor{black}{Using the notation of Eq.~(\ref{eq:Van_der-corput}), the $n$-th point of a two-dimensional Halton sequence is therefore $(V(n,2),V(n,3))$. Thus the same radical-inverse function $V(n,b)$ defines both the one-dimensional van der Corput sequence and each coordinate of the multidimensional Halton sequence.} Table~\ref{tab:halton} works through this digit-reversal explicitly for a few representative integers in bases $2$ and $3$.

\begin{table}[h]
\centering
\setlength{\tabcolsep}{4pt}
\begin{tabular}{c|l|l|c}
\hline\hline
$n$ & Numerical Value & Mirrored Value & \textcolor{black}{$V(n,b)$} \\
\hline
\multicolumn{4}{c}{Base(2) -- For 1D (X-axis)} \\
\hline
1 & $(1)_2 = 1\times2^0 = (1)_2$ & $V(1,2) = (0.1)_2 = 1\times 2^{-1}$ & 0.500 \\
2 & $(2)_2 = 1\times2^1 = (10)_2$ & $V(2,2) = (0.01)_2 = 1\times2^{-2}$ & 0.250 \\
6 & $(6)_2 = 1\times2^1 + 1\times2^2 = (110)_2$ & $V(6,2) = (0.011)_2 = 1\times2^{-2}+1\times2^{-3}$ & 0.375 \\
\hline
\multicolumn{4}{c}{Base(3) -- For 2D (Y-axis)} \\
\hline
4 & $(4)_3 = 1\times3^0 + 1\times3^1 = (11)_3$ & $V(4,3) = (0.11)_3 = 1\times3^{-1} + 1\times3^{-2}$ & 0.444 \\
7 & $(7)_3 = 1\times3^0 + 2\times3^1 = (21)_3$ & $V(7,3) = (0.12)_3 = 1\times3^{-1} + 2\times3^{-2}$ & 0.556 \\
\hline\hline
\end{tabular}
\caption{Illustration of the radical-inverse construction used in the Halton sequence.
The first coordinate is obtained from the van der Corput sequence in base $2$,
while the second coordinate is obtained from the corresponding sequence in
base $3$.}
\label{tab:halton}
\end{table}


Sobol sequences achieve low discrepancy through a different construction. Rather than assigning a distinct prime base to each dimension, they are built entirely in base $2$, with each coordinate generated from a set of direction numbers specific to that dimension. This base-$2$ construction tends to produce more uniformly structured point sets as dimensionality increases, which is why Sobol sequences are often preferred in higher-dimensional applications. We use the Halton and Sobol implementations provided by SciPy throughout this work, and for a more detailed treatment of their construction we refer the reader to Ref.~\citep{Dalal2008LDS}.


These quasirandom sequences are often \textit{scrambled}~\citep{OWEN1998466, 10.1145/3618307}, meaning that deterministic or random permutations are applied to their digits in a way that preserves low discrepancy. Scrambling breaks unwanted structural correlations, so that the sequences behave like random samples while retaining their uniform space-filling properties. \textcolor{black}{In our implementation the proposal sets are generated with the SciPy~\citep{Virtanen2020SciPy} Halton and Sobol generators. The sequences denoted Halton and Sobol are generated without scrambling, while Halton-S and Sobol-S use scrambling. The top-down template selection remains stochastic in all cases, so a deterministic LDS proposal set can still produce different bank realizations when the selection seed is changed.}

\begin{figure}[h]
    \centering
    \includegraphics[width=0.85\textwidth]{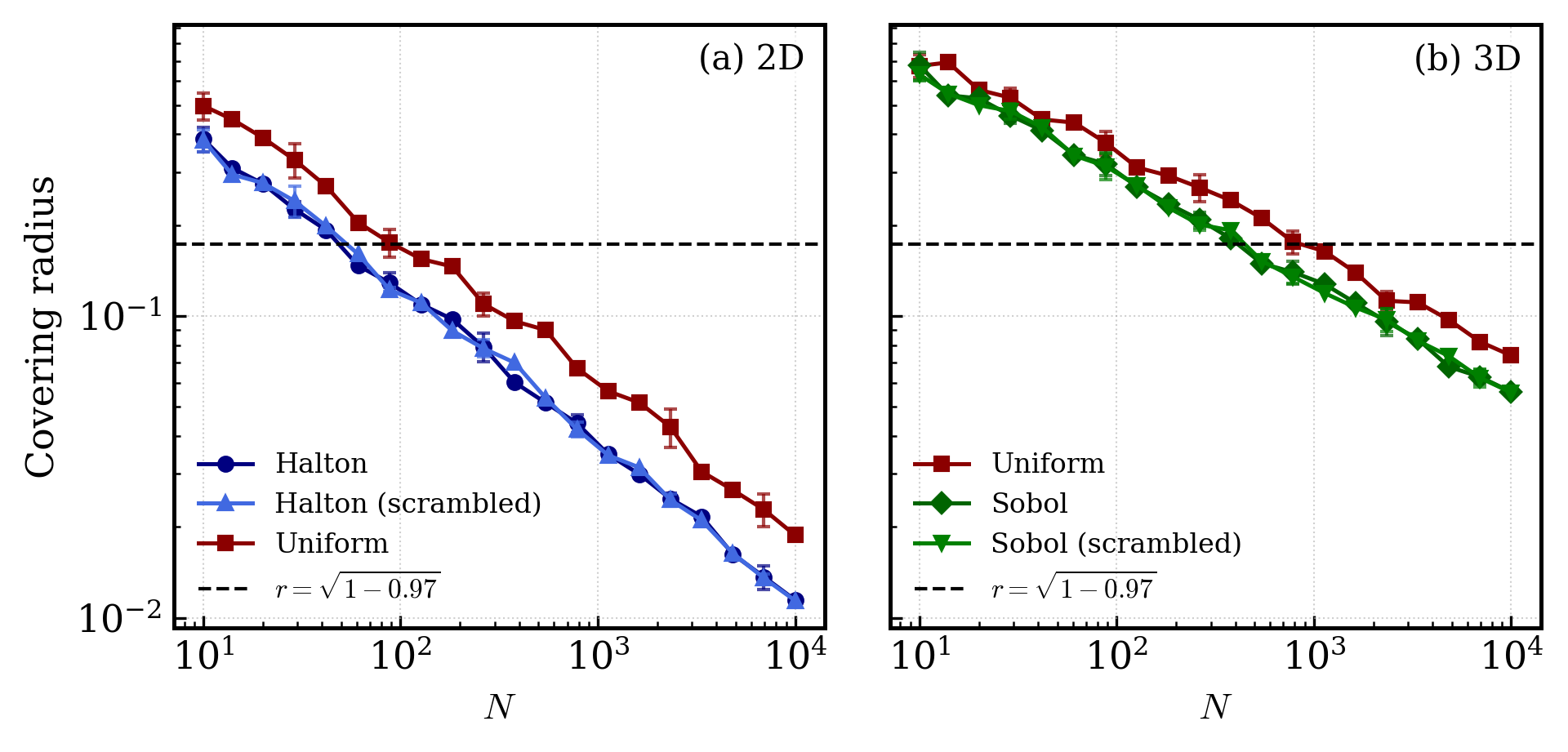}
    \caption{Scaling of the covering radius with $N$, for points placed in a unit space.  Error bars are the standard deviation over multiple realizations. Low-discrepancy sequences (LDS) cross the threshold $r=\sqrt{1-0.97}$, with $\sim 40\%$ fewer proposals compared to uniform sampling, demonstrating more efficient space-filling.}

    \label{fig:cqg}
\end{figure}


To illustrate the space-filling advantage of LDS over uniform random sampling, we generate $N$ points using uniform random, Halton, and Sobol sequences in $2$D and $3$D unit spaces, and track how coverage improves as $N$ increases (Fig.~\ref{fig:cqg}). We quantify coverage by the covering radius $r$, defined as the largest distance from any point in the space to its nearest sample point. Intuitively, $r$ is the radius of the largest empty ball that fits inside the sample, so it captures the worst-case gap in the coverage. In the context of gravitational-wave template bank generation, this quantity corresponds to the maximum mismatch, $r = \sqrt{1 - \mathcal{M}_{\mathrm{min}}}$. Operationally, for a bank of $N$ points, we evaluate $r$ by fixing $10{,}000$ random test points and taking the maximum (over test points) of the distance to the nearest bank point. We then track how rapidly $r$ falls below the threshold $r = \sqrt{1-0.97}$ as $N$ grows, motivated by the minimal match $\mathcal{M}_{\mathrm{min}} = 0.97$ standard in gravitational-wave template bank searches. As shown in Fig.~\ref{fig:cqg}, both Halton and Sobol sequences reach this threshold with approximately $40\%$ fewer points than uniform random sampling, confirming their more efficient space-filling in the dimensionalities relevant here.

\section{METHOD: TEMPLATE BANK CONSTRUCTION} 
\label{sec:Template bank construction}

\subsection{Metric on the signal manifold}
\label{sec:Parameter_space_metric}

Template bank construction may be formulated as a covering problem on the signal manifold generated by a family of gravitational-wave templates. 

\textcolor{black}{For the 2-D and 3-D template banks constructed using parameters given by Sets~I and II respectively in Table~\ref{tab:sets}, we employ the \IMRPhenomD~\citep{PhysRevD.93.044007} waveform model, whose frequency-domain representation may be written as}
$$
\tilde{h}(f;\vec{\lambda},\vec{\xi})
=
A(f;\vec{\lambda},\vec{\xi})
\,e^{-i \Psi(f;\vec{\lambda},\vec{\xi})},
\label{eq:waveform}
$$
where $\vec{\lambda}$ and $\vec{\xi}$ denote the intrinsic and extrinsic parameters, respectively; $A$ is the amplitude and $\Psi$ is the phase of the waveform. 
For aligned-spin \IMRPhenomD{} waveforms, the intrinsic parameter space consists of the component masses and the spin components aligned with the orbital angular momentum, i.e. 
${\vec{\lambda} = \{m_1,m_2,\chi_{1z},\chi_{2z}\}.}$


The local mismatch between nearby waveforms may be approximated using an intrinsic parameter-space metric $g_{ij}$. This metric is obtained by projecting the Fisher metric on the full signal manifold onto the intrinsic parameter subspace. Details of its construction are summarized in~\ref{sec:appendix_metric}.

\textcolor{black}{For the 4-dimensional bank whose parameters are summarized under Set~III of Table~\ref{tab:sets}, we evaluate the same projected Fisher metric $g$ numerically in the four coordinates $(\theta_0,\theta_3,\theta_{3s},e_{20}^2)$ using finite differences of the \texttt{TaylorF2Ecc} waveforms~\citep{Moore:2016suz}. The noise-weighted inner products entering the metric are evaluated from $20\,\mathrm{Hz}$ up to the Schwarzschild ISCO frequency.}

The mismatch neighbourhood of a template centred at $\vec{\lambda}_j$ is defined by
%
\begin{equation}
\Delta\vec{\lambda}^{\,T} \, 
g \, 
\Delta\vec{\lambda}
\leq
1-\mathcal{M}_{\rm min},
\label{eq:mismatch_ct}
\end{equation}
where 
${\Delta\vec{\lambda}=\vec{\lambda}-\vec{\lambda}_j}$ is the displacement to a neighbouring point in the intrinsic parameter space, and $\mathcal{M}_{\rm min}$ is the chosen minimal match.
%
Eq.~\ref{eq:mismatch_ct} defines the local region covered by a template: once a proposal is accepted as a template, any remaining proposal satisfying this condition is removed because its predicted mismatch with the accepted template is no larger than
$1-\mathcal M_{\min}$. This is a covering criterion and does not
impose a regular lattice spacing between accepted templates.
The template-bank construction can therefore be viewed as a sphere-covering problem on this effective parameter space, where the goal is to cover the target region using the smallest possible set of metric spheres of radius $\mathcal{D}_{\max} = \sqrt{1-\mathcal{M}_{\rm min}}.$


The use of the metric replaces expensive match evaluations with a local quadratic approximation to the waveform mismatch, thereby significantly reducing the computational cost of template-bank construction. 
In cases where the parameter-space metric is unavailable, distances between waveforms may instead be computed directly from \textcolor{black}{brute-force} overlap calculations.

For aligned-spin systems, it has been shown that the dominant spin effects may be captured by a single reduced-spin parameter
$\chi_r=\chi_s+\delta\,\chi_a-(76\eta/113)\chi_s, $
where $\chi_s$ and $\chi_a$ are the symmetric and antisymmetric spin combinations $(\chi_1\pm\chi_2)/2$, $\eta=m_1m_2/(m_1+m_2)^2$ is the symmetric mass ratio, and $\delta=(m_1-m_2)/(m_1+m_2)$ is the asymmetric mass ratio. This allows the four-dimensional intrinsic parameter space to be approximated by the three-dimensional manifold $\{m_1,m_2,\chi_r\}$ with minimal loss in effectualness \citep{PhysRevD.89.084041}.

This parameter space may be further reparametrized using the dimensionless chirp-time coordinates $(\theta_0,\theta_3,\theta_{3s})$. 
The transformation from $(m_1,m_2,\chi_r)$ to chirp-time coordinates yields a parametrization in which the metric components vary more slowly across the parameter space, making it particularly well suited for metric-based template placement.
The dimensionless chirp-time coordinates are defined by
\begin{equation}
\theta_{0} =
\frac{5}{2^{1/3}}
\left(
\frac{1}{16\pi f_{0}m\eta^{3/5}}
\right)^{5/3},
\qquad
\theta_{3} =
\left(
\frac{16\pi^{5}\theta_{0}^{2}}
{25\eta^{3}}
\right)^{1/5},
\qquad
\theta_{3s} =
\frac{113}{48\pi}\chi_r\theta_{3}.
\label{eq:Co-ordinates}
\end{equation}
where \(m\) denotes the total mass of the binary, and \(f_{0}\) is the lower frequency cutoff.
\textcolor{black}{For Sets~I and II, we construct 2D template banks over the intrinsic parameter space $(\theta_0,\theta_3)$ and 3D template banks over $(\theta_0,\theta_3,\theta_{3s})$, respectively. In the 2D case, the spin effects are neglected by setting $\theta_{3s}=0$.}


\subsection{Template placement algorithm}
\label{sec:Template placement algorithm}

Several techniques have been proposed for the placement of template banks in the intrinsic parameter space of compact-binary signals \citep{PhysRevD.80.104014, PhysRevD.108.042003, Kacanja_2024, PhysRevD.86.084017, PhysRevD.89.084041, PhysRevD.93.124007}. As discussed in Sec.~\ref{sec:intro}, the stochastic template placement algorithm may be implemented in either a bottom-up or a top-down manner; we use the top-down approach, which begins from an oversampled proposal set and iteratively removes proposals already covered by a selected template, leaving the surviving selected points as the final bank.

We first generate a proposal set $\mathcal{U}$ in the intrinsic parameter space. In the standard stochastic approach this set is drawn from a uniform random distribution and oversampled to ensure adequate coverage; here we additionally compare this choice against proposal sets generated from low-discrepancy sequences (Halton and Sobol), which cover the parameter space more uniformly and can yield banks of comparable size and coverage from fewer initial proposals. To construct $\mathcal{U}$, we determine the allowed ranges of the chirp-time coordinates from the mass and spin ranges in Table~\ref{tab:sets}, sample points over these coordinates, and map them back to physical masses and spins; only points satisfying the physical constraints and bounds of Table~\ref{tab:sets} are retained. Each point in $\mathcal{U}$ carries a Boolean active status, initially true for all proposals, and we initialize an empty template list $\mathcal{T}$. A Euclidean KD-tree is built once, using all points in $\mathcal{U}$, before the placement loop begins; points are never removed from the tree itself, only their active status is updated, thereby avoiding the cost of rebuilding it after every accepted template. \textcolor{black}{A KD-tree is a binary spatial index that partitions the coordinate space and permits fast range queries. Here ``Euclidean'' refers only to the ordinary coordinate-space distance used for this conservative broad query; the final covering decision is always made with the local parameter-space metric.}

\begin{figure}[t]
\centering
\includegraphics[width=0.95\textwidth]{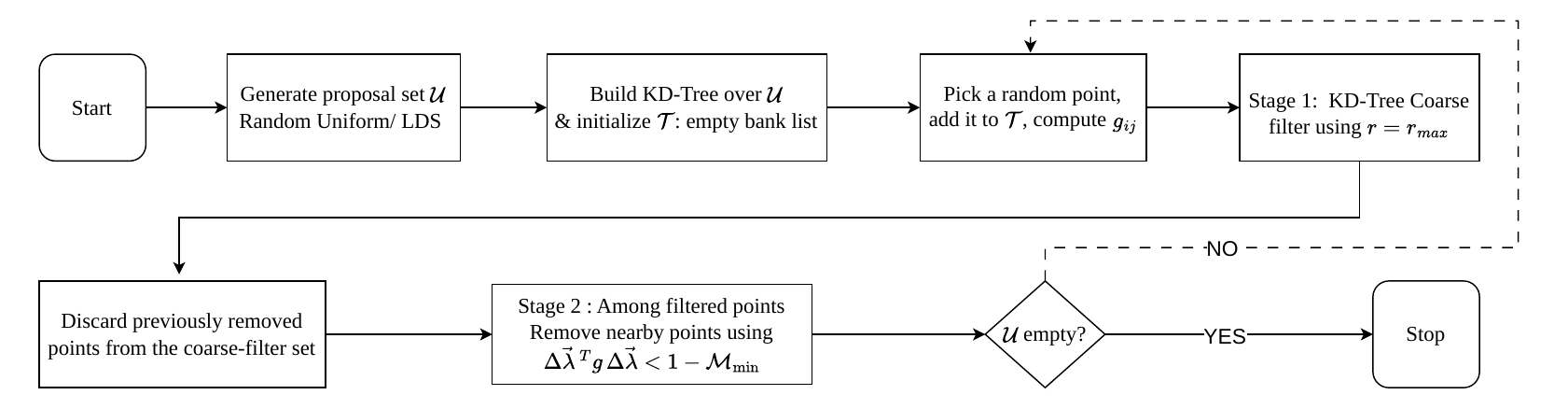}
\caption{
Flowchart for the top-down stochastic template bank construction algorithm. A proposal set $\mathcal{U}$ is generated once, using either traditional uniform random sampling or the LDS approach (proposed in this work), and a KD-tree is built over these points. At each iteration, a randomly chosen active proposal is promoted to the template list $\mathcal{T}$. Candidate neighbors of this proposal are first identified with a KD-tree broad query, then tested against the exact local metric condition. Covered proposals are marked inactive, and the process continues until no active proposals remain.
}
\label{fig:flowchart}
\end{figure}

At each iteration, one active proposal $\lambda_j\in\mathcal{U}$ is chosen at random and promoted to $\mathcal{T}$, and the local parameter-space metric $g(\lambda_j)$ is computed at this point. A neighbouring proposal $\lambda$ is covered by the template if the local quadratic mismatch condition, Eq.~(\ref{eq:mismatch_ct}), is satisfied. 

Diagonalizing the local metric,
$
g(\lambda_j)=Q\, \Lambda \, Q^T ,
$
where the columns of $Q$ are the eigenvectors $q_k$ and $\Lambda={\rm diag}(\alpha_1,\ldots,\alpha_n)$, Eq.~\eqref{eq:mismatch_ct} becomes
\begin{equation}
\sum_{k=1}^{n}\alpha_k \, u_k^2 \leq 1-\mathcal{M}_{\min},
\end{equation}
in the metric eigenbasis, where $u_k=\Delta\lambda\cdot q_k$.
The mismatch neighbourhood is thus an ellipsoid with principal axes aligned along $q_k$ and semi-axis lengths
\begin{equation}
a_k = \sqrt{ \frac{1-\mathcal{M}_{\min}}{\alpha_k} } .
\end{equation}
\textcolor{black}{For the calculations reported here we use a fixed safety factor $\epsilon=0.005$ in the coarse search, $\mathcal{M}_{\min}$ is effectively reduced to  $\mathcal{M}_{\min}-\epsilon$, giving correspondingly enlarged semi-axes $a_k\rightarrow a_k^{(\epsilon)}$. The pad is used only to make the coarse candidate search conservative; increasing it enlarges the candidate set and hence the coarse-search cost, but does not change the final covering criterion, which always uses the exact threshold in Eq.~\eqref{eq:mismatch_ct}.}

Testing Eq.~(\ref{eq:mismatch_ct}) against every active proposal in $\mathcal{U}$ is inefficient, so we first perform a coarse Euclidean search using the KD-tree. Let $\alpha_{\min}$ denote the minimum eigenvalue of $g(\lambda_j)$. Since
\begin{equation}
\alpha_{\min} \, \|\lambda - \lambda_j\|^2 \leq (\lambda - \lambda_j)^T \, g(\lambda_j) \, (\lambda - \lambda_j),
\end{equation}
every point satisfying Eq.~(\ref{eq:mismatch_ct}) must lie inside the Euclidean ball
\begin{equation}
\|(\lambda - \lambda_j)\| \leq r_{\max},
\label{eq:euclidean_ball}
\end{equation}
where
\begin{equation}
\textcolor{black}{r_{\max} = \sqrt{\frac{1-(\mathcal{M}_{\min}-\epsilon)}{\alpha_{\min}}}
= \sqrt{\frac{1-\mathcal{M}_{\min}+\epsilon}{\alpha_{\min}}}.}
\label{eq:r_max}
\end{equation}
\textcolor{black}{This padded Euclidean ball of radius $r_{\max}$ is a conservative bounding region for the metric ellipsoid. Querying the KD-tree with this radius around $\lambda_j$ returns a coarse candidate set containing all proposals that could possibly satisfy the exact covering condition, along with possibly many points outside the true ellipsoid when the local metric is anisotropic.} 

Since the query operates on the original set $\mathcal{U}$, returned candidates may include points already marked inactive in a previous iteration; these are discarded, and the remaining active candidates are tested against the exact condition, Eq.~(\ref{eq:mismatch_ct}). 
All active proposals satisfying this condition, together with
$\lambda_j$ itself, are marked inactive. The algorithm then selects
another random active point and repeats, terminating when no active
proposals remain. At termination, every point in $\mathcal{U}$ is
either included in the template bank or lies within the metric
neighbourhood of a selected template. The final bank is the list
$\mathcal{T}$.
%

\begin{figure}[h!]
\centering
\includegraphics[width=0.85\textwidth]{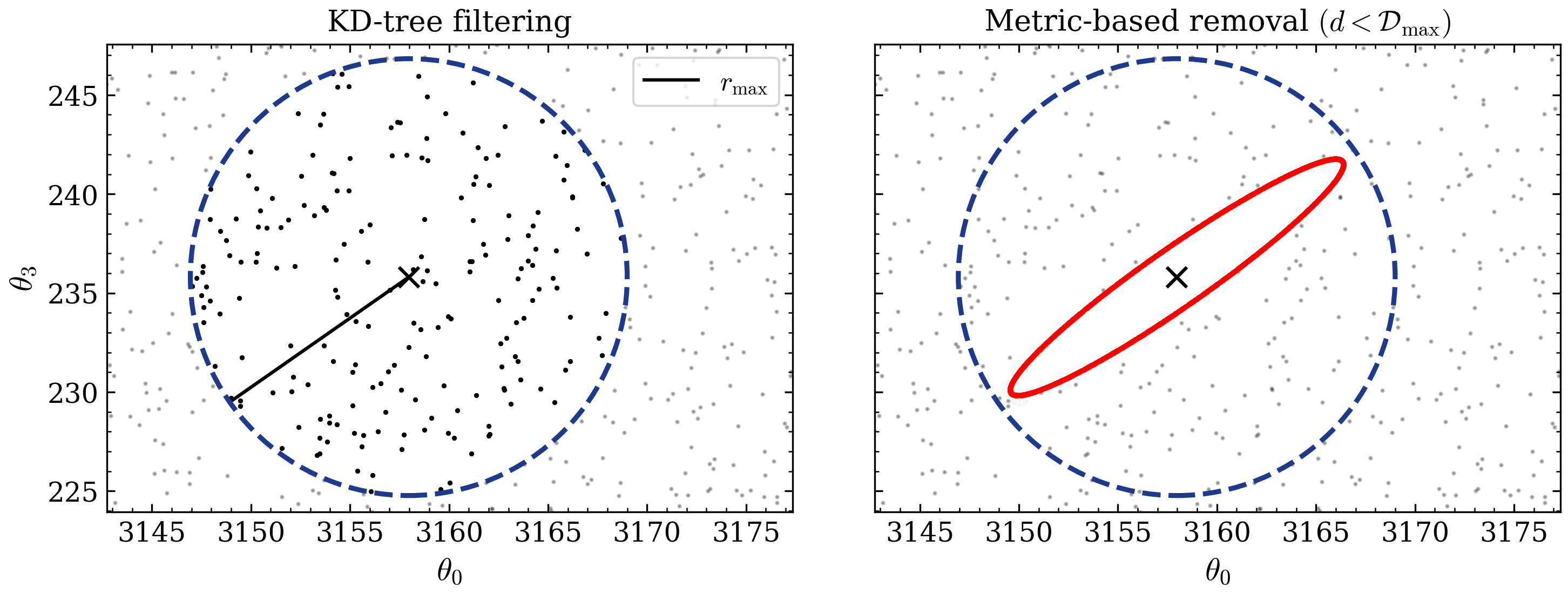}
\caption{
Illustration of the two-step point-deletion algorithm. The KD-tree filtering step, shown in the left panel, identifies proposals within the conservative bounding radius $r_{\max}$ of the reference template $\lambda_{\mathrm{ref}}$; this radius is defined in Eq.~(\ref{eq:r_max}). The surviving active proposals are then tested against the metric criterion, shown in the right panel. Points satisfying Eq.~(\ref{eq:mismatch_ct}) are removed from the active proposal set $\mathcal{U}$.
}
\label{fig:Illustration Algorithm}
\end{figure}

This two-stage deletion procedure is illustrated in Fig.~\ref{fig:flowchart} and Fig.~\ref{fig:Illustration Algorithm}: a KD-tree broad query reduces the candidate set via a conservative Euclidean bound, followed by an exact metric cut that enforces the true local mismatch criterion.
\textcolor{black}{We refer to the candidate-search and deletion operations associated with each accepted template as one \emph{covering step}. In the baseline algorithm, this consists of the KD-tree broad query followed by the exact metric cut. The covering stage of the full bank construction is the accumulation of these operations over all accepted templates.}

Once a template bank $\mathcal{T}$ has been constructed, its effectiveness is assessed using random injections drawn from the same parameter space. For each injection we compute the fitting factor \citep{PhysRevD.52.605}, the maximum overlap between an arbitrary signal $h_a$ and the templates $h$ in the bank:
\begin{equation}
\mathcal{FF}(h_a) = \max_{h\in\mathcal{T}} \langle h_a \mid h \rangle .
\label{eq Factor}
\end{equation}
Bank performance is evaluated via the cumulative distribution function of the fitting factors, together with the recovery fraction $\mathcal R_{0.97}$, defined as the percentage of injections with $\mathcal{FF}\geq 0.97$.

\subsection{Slab-wise coarse filtering for anisotropic metrics}
\label{sec:slabwise_filtering}

\begin{figure}[t]
    \centering
    \includegraphics[width=0.75\textwidth]{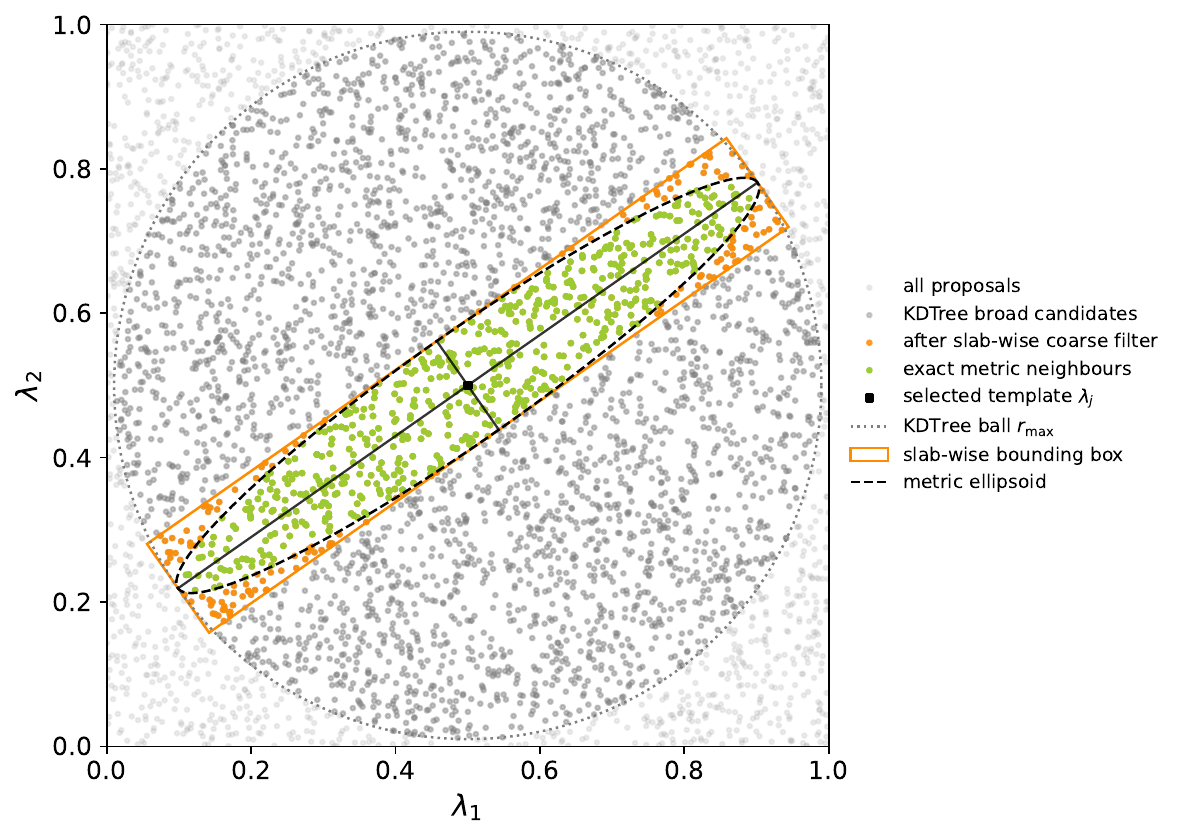}
    \caption{
    Illustration of the slab-wise coarse filtering step in two dimensions. The
    selected template is shown at $\lambda_j$, and
    $\Delta\lambda=\lambda-\lambda_j$. The KD-tree broad query first returns
    all proposals satisfying $\|\Delta\lambda\|\leq a_{\max}^{(\epsilon)}$. The slab-wise coarse filter then keeps
    only candidates inside the oriented bounding box defined by
    $|\Delta\lambda\cdot q_k|\leq a_k^{(\epsilon)}$ for each metric eigenvector
    $q_k$. Finally, the exact metric cut selects the proposals satisfying
    $\Delta\lambda^T g(\lambda_j)\Delta\lambda
    \leq 1-\mathcal{M}_{\min}$.
    }
    \label{fig:slab_cut_demo_2d}
\end{figure}

The induced metric $g$ on the signal manifold is often highly anisotropic; the latter quantified by the condition number
$
    \kappa(g)
    =
    {\alpha_{\max}}/{\alpha_{\min}},
$
where $\alpha_{\max}$ and $\alpha_{\min}$ are the largest and smallest eigenvalues of $g(\lambda_j)$. For $\kappa \gg 1$, the metric ellipsoid is highly elongated. As a result, the conservative KD-tree selection around a chosen template over-approximates the true mismatch ellipsoid and returns a very large number of proposals that are subsequently rejected by the exact metric cut (see Fig.~\ref{fig:slab_cut_demo_2d}).

We therefore introduce an intermediate slab-wise coarse filtering step using the oriented bounding box of the metric
ellipsoid as:
\begin{equation}
    |\Delta\lambda\cdot q_k|
    \leq
    a_k^{(\epsilon)},
    \qquad k=1,\ldots,n .
    \label{eq:slab_cut}
\end{equation}
Each inequality defines a slab bounded by two parallel hyperplanes normal to
$q_k$. The oriented box is the intersection of these slabs. In practice, the
slabs are applied sequentially, starting with the smallest semi-axis, so that
the most restrictive directions reject candidates first. Since the KD-tree
ball already enforces $\|\Delta\lambda\|\leq a_{\max}^{(\epsilon)}$, the
widest slab is automatically satisfied by all KD-tree candidates and is
therefore skipped while applying the slab cut.

\textcolor{black}{When the slab-wise pre-filter is used, each covering step proceeds in three stages.}
First, the KD-tree is used to obtain a conservative Euclidean candidate set
around the selected proposal. Second, this candidate set is further reduced
using the slab-wise coarse filter, which keeps only points lying inside the
oriented bounding box of the metric ellipsoid. Finally, the remaining
candidates are tested using the exact quadratic metric condition. It is important to note that the
slab-wise step is only a pre-filter to reduce the computational cost of
each covering step; it does not change the active status of
any proposal. A proposal is marked inactive only after it satisfies the final
metric cut.
\textcolor{black}{The 2D and 3D banks presented in Sec.~\ref{sec:Results} use the baseline two-stage procedure, consisting of the KD-tree broad query followed directly by the exact metric cut; the slab-wise pre-filter is not used in the corresponding bank-construction timings.}

Figure~\ref{fig:slab_cut_demo_2d} illustrates the KD-tree ball, slab-wise
coarse filter, and exact metric neighbors in two dimensions.

The advantage of the slab filter is most apparent from candidate-count
scaling. For a locally uniform proposal density $\rho$, the expected number of
KD-tree candidates scales as the volume of the Euclidean bounding ball,
$
    N_{\rm KD}
    \sim
    \rho V_n a_{\max}^n ,
$
where $V_n$ is the volume of the unit $n$-ball. On the other hand, the expected
number of points inside the metric ellipsoid scales as
$
    N_{\rm exact}
    \sim
    \rho V_n \prod_{k=1}^{n} a_k .
$
Thus
\begin{equation}
    \frac{N_{\rm KD}}{N_{\rm exact}}
    \sim
    \frac{a_{\max}^{n}}
    {\prod_{k=1}^{n} a_k}
    \sim 
    \sqrt{
    \frac{\prod_{k=1}^{n}\alpha_k}
    {\alpha_{\min}^{\,n}}
    } .
    \label{eq:kdtree_exact_scaling}
\end{equation}
This ratio grows rapidly with both dimension and anisotropy. The Euclidean
KD-tree ball can therefore contain orders of magnitude more proposals than the
actual metric neighbourhood.

By contrast, the slab-wise step replaces the Euclidean ball by an oriented box
adapted to the metric eigenvectors. The box volume scales as
$
    V_{\rm box}=2^n\prod_{k=1}^{n}a_k ,
$
so that
\begin{equation}
    \frac{N_{\rm box}}{N_{\rm exact}}
    \sim
    \frac{2^n}{V_n}.
\end{equation}
This ratio is independent of metric anisotropy. The slab-wise filter in
Eq.~\eqref{eq:slab_cut} is therefore better tailored to the geometry of the
metric ellipsoid.

\begin{table}[h]
\centering
\begin{tabular}{c c c c c c}
\hline\hline
$n$
& $N_{\rm KD}$
& $N_{\rm slab}$
& $N_{\rm exact}$
& $N_{\rm KD}/N_{\rm slab}$
& Speed-up \\
\hline
$2$ & $2.28\times10^{5}$ & $3.09\times10^{3}$ & $2.49\times10^{3}$
& $73.97$ & $0.90$ \\
$3$ & $7.18\times10^{4}$ & $1.54\times10^{2}$ & $85.68$
& $465.02$ & $1.10$ \\
$4$ & $2.01\times10^{4}$ & $8.72$ & $3.58$
& $2303.05$ & $1.23$ \\
$5$ & $4.92\times10^{3}$ & $1.35$ & $1.03$
& $3641.32$ & $1.27$ \\
$6$ & $1.14\times10^{3}$ & $1.04$ & $1.00$
& $1096.61$ & $1.33$ \\
\hline \hline
\end{tabular}
\caption{
Local-query benchmark of the slab-wise coarse filtering step. The reported
numbers are averages over random query points in an $n$-dimensional unit
hypercube. Here $N_{\rm KD}$ is the average number of candidates returned by
the KD-tree broad query, $N_{\rm slab}$ is the average number remaining after
slab-wise coarse filtering, and $N_{\rm exact}$ is the average number
satisfying the final metric cut. The slab-wise and direct pipelines were
verified to return the same final set of metric neighbours. The speed-up is
the ratio of the runtime of the baseline ``KD-tree + exact-cut'' method to
that of the ``KD-tree + slab-cut + exact-cut'' method.
}
\label{tab:slabwise_benchmark}
\end{table}
We tested the slab-wise coarse filter using a controlled local-query
benchmark. We generated $N=1.5\times 10^6$ proposal points uniformly in the
unit hypercube $[0,1]^n$ for $n=2$--$6$ and built a Euclidean KD-tree on each
fixed proposal set, then selected $100$ random query points. For each query
point $\lambda_j$ we assigned a synthetic positive-definite metric
$
    g(\lambda_j)=Q\,\Lambda \,Q^T ,
$
where $Q$ is a random orthogonal matrix and the eigenvalues in $\Lambda$ are
logarithmically spaced between $1$ and $10^4$, giving a strongly anisotropic,
randomly oriented metric ellipsoid with condition number $\kappa=10^4$. The
KD-tree broad-query radius was fixed to $r_{\max}=0.25$. For each query we
compared (a) the baseline consisting of the KD-tree broad query~(Eq.~\eqref{eq:euclidean_ball})
followed directly by the exact metric cut~(Eq.~\eqref{eq:mismatch_ct})  against (b) 
the slab-wise method, which applies the slab-wise coarse
filter~(Eq.~\eqref{eq:slab_cut}) before the exact metric cut. We verified that
both methods return the same final set of metric neighbours.

Results, averaged over the $100$ queries, are summarized in
Table~\ref{tab:slabwise_benchmark}. The slab-wise filter sharply reduces the
number of candidates passed to the exact metric cut, and thus the number of
full quadratic-form evaluations, with the benefit growing in higher
dimensions: for $n>3$ it cuts the local covering-step time by about
$20$--$30\%$ relative to the baseline.

\textcolor{black}{
To test the slab-wise pre-filter in a full CBC bank construction, we construct a four-dimensional eccentric equal-spin toy bank using the \texttt{TaylorF2Ecc} waveform model~\citep{Moore:2016suz, lalsuite} and the \aLIGOZeroDetHighPower{}~\citep{LIGO-T0900288} noise power spectral density (PSD).
The intrinsic coordinates are $(\theta_0,\theta_3,\theta_{3s},e_{20}^2)$.  The component masses $m_1,m_2\in[5,20]M_\odot$ subject to $m_1\ge m_2$, equal aligned spins $\chi_{1z}=\chi_{2z}\equiv\chi\in[-0.5,0.5]$, and eccentricity $e_{20}\in[0,0.15]$ defined at $20\,\mathrm{Hz}$. 
We use $e_{20}^2$ rather than $e_{20}$ because the leading eccentric correction to the \texttt{TaylorF2Ecc} phase is quadratic in eccentricity, making $e_{20}^2$ the natural local coordinate near the circular limit.
The local 4D metric is obtained from the \texttt{TaylorF2Ecc} Fisher matrix by finite-differencing the waveform amplitude and phase with respect to the four intrinsic coordinates, including the coalescence-time and phase derivatives analytically, and projecting out these two extrinsic directions. The metric inner products are evaluated from $f_{\rm low}=20\,\mathrm{Hz}$ up to the Schwarzschild ISCO frequency. We use $\mathcal M_{\min}=0.97$. For the ablation below, the proposal-set size is fixed at $|\mathcal U|=5\times10^5$.}

\textcolor{black}{
The 4D toy bank also exhibits a reduction in the required proposal-set size with LDS sampling.
Using a common set of $50{,}000$ injections, a random-uniform proposal set with $|\mathcal U|=6\times10^5$ gives $\mathcal R_{0.97}=0.98466$, while a Sobol proposal set with $|\mathcal U|=5\times10^5$ gives $\mathcal R_{0.97}=0.98472$.
Thus, at essentially identical recovery, Sobol sampling requires $16.7\%$ fewer proposals. %
A local monotonic cubic Hermite interpolation~\citep{FritschButland1984} of the recovery curves gives an equivalent Sobol size of $4.986\times10^5$, corresponding to a $\sim 17\%$ reduction, consistent with the direct matched-recovery comparison. These numbers quantify the LDS reduction in proposal-set size.}

\textcolor{black}{The slab-filter ablation addresses a separate question: to isolate the computational effect of the slab pre-filter from the reduction in $|\mathcal U|$ due to LDS sampling, we deliberately fix $|\mathcal U|=5\times10^5$ for both Random and Sobol proposals and vary only the single-slab pre-filter while keeping all other settings unchanged. Thus, the ablation measures the slab contribution alone and does not mix it with the approximately $17\%$ LDS proposal-set reduction.}

\begin{table}[t]
\centering
\color{black}
\setlength{\tabcolsep}{5pt}
\renewcommand{\arraystretch}{1.015}
\begin{tabular}{l c r r r r}
\hline\hline
Sampling & Slab & $N_{\rm temp}$ &
$\langle N_{\rm eval}\rangle$ &
$t_{\rm cover}$ (s) & $t_{\rm total}$ (s) \\
\hline
Random & off & 19,442 & $8.63\times10^4$ & 162.49 & 493.29 \\
Random & on  & 19,442 & $4.01\times10^2$ & 55.65 & 386.09 \\ \hline
Sobol  & off & 19,746 & $8.55\times10^4$ & 163.44 & 500.77 \\
Sobol  & on  & 19,746 & $3.99\times10^2$ & 56.16 & 393.37 \\
\hline\hline
\end{tabular}
\caption{Ablation of the single-slab pre-filter for the 4D eccentric equal-spin toy bank at the common proposal-set size $|\mathcal U|=5\times10^5$. The same value of $|\mathcal U|$ is used for Random and Sobol so that the slab-filter gain is measured independently of the LDS reduction in proposal-set size. Here $\langle N_{\rm eval}\rangle$ is the mean number of candidates passed to the exact metric test per covering step. For a given proposal set, switching the single-slab filter on or off gives the same final bank.}
\label{tab:4d_slab_ablation}
\end{table}

\textcolor{black}{For the 4D bank, the four coordinates have very different numerical ranges, with $e_{20}^2$ spanning a much smaller interval than the chirp-time coordinates. Because the Euclidean distance used by the KD-tree depends on these coordinate scales, we first map each coordinate to the unit interval,
$x^i=(\lambda^i-\lambda^i_{\min})/(\lambda^i_{\max}-\lambda^i_{\min})$.
Writing $S_{ii}=\lambda^i_{\max}-\lambda^i_{\min}$, so that $\Delta\lambda=S\,\Delta x$, the metric in the normalized coordinates is $g_x=S^T g S$. Hence
$\Delta x^T g_x\Delta x=\Delta\lambda^T g\Delta\lambda$, so the quadratic mismatch is unchanged. This normalization is only a numerical reparametrization that places the coordinates on comparable scales for the KD-tree and slab operations; it does not alter the physical covering criterion. In these unit-box coordinates the local 4D metric is strongly anisotropic: its condition number has a median of approximately $9.5\times10^7$, with the upper end of the sampled range reaching $\sim1.7\times10^9$. We therefore use only the narrowest metric-eigenvector slab as the coarse pre-filter in the full 4D bank construction. This single slab already rejects more than $99\%$ of the KD-tree candidates. Once this reduction has been made, the remaining exact quadratic-form test is inexpensive, so applying the other slabs gives little additional benefit compared with their filtering overhead. The ablation below therefore compares the baseline KD-tree broad query followed by the exact metric cut against the KD-tree broad query followed by the single narrowest slab and then the exact metric cut.}

\textcolor{black}{For this ablation we record the final template count
$N_{\rm temp}$, the mean number of candidates passed to the exact quadratic
metric test per covering step, $\langle N_{\rm eval}\rangle$, the
covering-stage time $t_{\rm cover}$, and the total bank-construction time
$t_{\rm total}$. In the normalized coordinates this exact test is
$\Delta x^T g_x \Delta x \leq 1-\mathcal M_{\min}$, which is equivalent to
Eq.~\eqref{eq:mismatch_ct} under the coordinate transformation. A proposal may
be tested in several covering steps before it is removed, so
$\langle N_{\rm eval}\rangle$ counts metric evaluations per accepted template
rather than unique proposals.}




\textcolor{black}{At fixed $|\mathcal U|=5\times10^5$, enabling the single-slab
pre-filter leaves the final Random and Sobol banks unchanged, while reducing
$\langle N_{\rm eval}\rangle$ by $99.53\%$. The covering-stage runtime is
reduced by factors of $2.92$ and $2.91$ for Random and Sobol sampling,
respectively. The corresponding total bank-construction speed-ups are
$1.28$ and $1.27$, since metric evaluation remains a substantial part of the
total runtime.}

\textcolor{black}{The slab-wise filter can also reduce the cost of
fitting-factor calculations by restricting expensive waveform overlaps to
a small set of metric-selected candidate templates.}

\section{RESULTS}
\label{sec:Results}
\textcolor{black}{In this section, we present results for template banks constructed in 2D, 3D, and 4D using random-uniform and low-discrepancy-sequence (LDS) proposal sets. Table~\ref{tab:sets} summarizes the three parameter sets. Sets~I and II use the \IMRPhenomD{} waveform model, while Set~III uses \texttt{TaylorF2Ecc}; all three use the \aLIGOZeroDetHighPower{}~\citep{LIGO-T0900288} PSD.} 

\begin{table*}[b]
\begin{tabular}{p{0.25\textwidth} | p{0.21\textwidth} | p{0.21\textwidth} | p{0.21\textwidth}}
\hline \hline
Bank Parameter & Set-I(2D) & Set-II(3D) & \textcolor{black}{Set-III(4D)} \\
\hline
Waveform model & \IMRPhenomD{} & \IMRPhenomD{} & \textcolor{black}{\texttt{TaylorF2Ecc}} \\
Lower cut-off frequency $f_{\rm low}$ & $20\,\mathrm{Hz}$ & $20\,\mathrm{Hz}$ & \textcolor{black}{$20\,\mathrm{Hz}$} \\
Mass of first object $m_{1}$ & $[1,20]\,M_{\odot}$ & $[1,20]\,M_{\odot}$ & \textcolor{black}{$[5,20]\,M_{\odot}$} \\
Mass of second object $m_{2}$ & $[1,3]\,M_{\odot}$ & $[1,3]\,M_{\odot}$ & \textcolor{black}{$[5,20]\,M_{\odot}$} \\
Spin of first object $\chi_{1}$ & $--$ & $[-0.98,0.98]$ & \textcolor{black}{$[-0.5,0.5]$} \\
Spin of second object $\chi_{2}$ & $--$ & $[-0.04,0.04]$ & \textcolor{black}{$\chi_2=\chi_1$} \\
\textcolor{black}{Eccentricity $e_{20}$} & \textcolor{black}{$--$} & \textcolor{black}{$--$} & \textcolor{black}{$[0,0.15]$} \\
Minimal Match $M_{\min}$ & $0.97$ & $0.97$ & \textcolor{black}{$0.97$} \\
\hline \hline
\end{tabular}
\caption{Different sets of parameters used to generate the banks. The parameter space for Set~II is constructed by applying the NSBH 
mass boundary criterion, wherein objects having individual masses at 
or below $3\,M_{\odot}$ are treated as neutron stars, with their 
dimensionless spin magnitudes restricted to a range of $\pm 0.04$. \textcolor{black}{All three banks use the \aLIGOZeroDetHighPower{}~\citep{LIGO-T0900288} noise PSD. For Set~III, $m_1\geq m_2$, the aligned component spins are constrained to be equal, $\chi_1=\chi_2$, and $e_{20}$ is defined at $20\,\mathrm{Hz}$, with $f_{\rm ecc}=20\,\mathrm{Hz}$.}}
\label{tab:sets}
\end{table*}

\textcolor{black}{
The final template count in a top-down stochastic algorithm can depend on
$|\mathcal{U}|$ when the proposal set is too sparse to sample the parameter
space adequately. As $|\mathcal{U}|$ increases, the bank size approaches a
stable value, while smaller proposal sets may leave parts of the parameter
space poorly sampled and therefore give both fewer templates and a lower
recovery fraction $\mathcal R_{0.97}$. We therefore study
$\mathcal R_{0.97}$ as a function of $|\mathcal{U}|$ directly.}

\textcolor{black}{
To quantify the LDS advantage, we match the recovery attained by the
random-uniform bank to the corresponding LDS recovery curve and determine the
equivalent LDS proposal-set size by monotonic interpolation between the
measured points. For a selected random-uniform recovery
$R_\star=\mathcal R_{0.97}^{\rm Rand}(|\mathcal U|_{\rm Rand})$, the equivalent
LDS size $|\mathcal U|_{\rm LDS}$ satisfies
$\mathcal R_{0.97}^{\rm LDS}(|\mathcal U|_{\rm LDS})=R_\star$. The absolute
proposal-set reduction is
$\Delta|\mathcal U|=|\mathcal U|_{\rm Rand}-|\mathcal U|_{\rm LDS}$, and we
define the fractional LDS advantage by}
\begin{equation}
\textcolor{black}{
\Delta_{\rm LDS}(\%)
=
100\left(
1-\frac{|\mathcal U|_{\rm LDS}}
{|\mathcal U|_{\rm Rand}}
\right).
}
\label{eq:lds_advantage}
\end{equation}

One practical way to obtain an initial scale for $|\mathcal{U}|$ is to estimate the total volume of the parameter space, as discussed in~\citet{PhysRevD.95.104045}. 
\textcolor{black}{In the present comparison, the LDS reduction is instead determined directly from the measured recovery curves at matched $\mathcal R_{0.97}$.}

\begin{figure}[t]
    \centering
    \includegraphics[width=\textwidth]{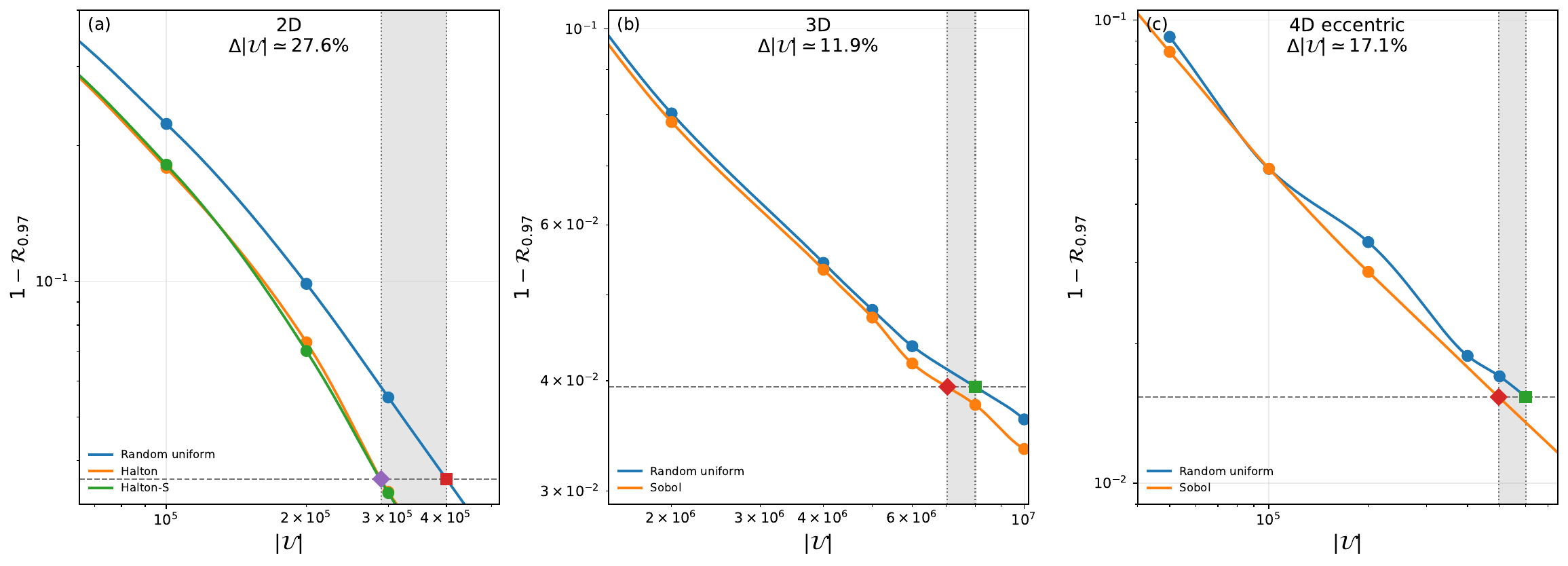}
    \caption{\textcolor{black}{Unrecovered fraction $1-\mathcal R_{0.97}$ as a
    function of the proposal-set size $|\mathcal U|$ for the 2D, 3D, and 4D
    banks. Points show the measured values and solid curves are the monotonic
    PCHIP interpolants~\citep{FritschButland1984}, plotted here as $1-\mathcal R_{0.97}$. Smaller values
    correspond to higher recovery. The 2D panel compares random-uniform,
    Halton, and scrambled-Halton sampling; the 3D panel compares random-uniform
    and Sobol sampling; and the 4D panel shows the corresponding comparison for
    the eccentric equal-spin toy bank. The horizontal dashed line marks the
    selected random-uniform recovery level, expressed as $1-R_\star$, while
    the dotted vertical lines indicate the random-uniform proposal size and the
    equivalent LDS proposal size at matched recovery. The shaded region
    therefore represents the absolute reduction $\Delta|\mathcal U|$, while
    the corresponding fractional LDS advantage is defined by
    Eq.~\eqref{eq:lds_advantage}. The matched-recovery reductions shown here
    are approximately $28\%$, $12\%$, and $17\%$ in 2D, 3D, and 4D,
    respectively.}}
    \label{fig:R097_vs_U}
\end{figure}

\textcolor{black}{
The recovery-fraction curves shown in Fig.~\ref{fig:R097_vs_U} provide a
direct estimate of the reduction in $|\mathcal U|$ at matched recovery.
Using Eq.~\eqref{eq:lds_advantage}, the recovery of the 2D random-uniform
bank at $|\mathcal U|=4\times10^5$ is matched on the monotonic Halton
curve at $|\mathcal U|\simeq2.9\times10^5$, corresponding to a reduction
of about $28\%$. In 3D, the random-uniform bank at
$|\mathcal U|=8\times10^6$ has $\mathcal R_{0.97}=0.96062$; the same
recovery is obtained on the Sobol curve at
$|\mathcal U|=7.045\times10^6$, corresponding to a reduction of
$11.94\%$. At $|\mathcal U|=10^7$, the corresponding reduction is
$13.17\%$. We therefore quote an approximately $12\%$ reduction in
proposal-set size for the 3D case.
}

\textcolor{black}{
To make the small differences between Random and LDS sampling more visible
in the 3D and 4D cases, we plot the unrecovered fraction
$1-\mathcal R_{0.97}$ on a logarithmic scale.
}

A reduction in $|\mathcal U|$ at fixed recovery reduces the memory and proposal-processing costs of bank generation.
In addition, the memory footprint of the proposal set and the associated KD-tree~\cite{Bentley1975} search structures also scale with $|\mathcal{U}|$, making reductions in $|\mathcal{U}|$ particularly valuable for higher-dimensional searches.

\textcolor{black}{We next test whether the proposal-set sizes selected from the recovery curves give consistent results across independent bank realizations. For Sets~I and II, we construct 10 independent banks for each sampling method at the fixed values of $|\mathcal U|$ listed in Table~\ref{tab:results}.}
The corresponding recovery fractions are evaluated using a common set of $10^5$ random injections drawn from the same parameter space.

\begin{table}[h]
\centering
\setlength{\tabcolsep}{5pt}
\renewcommand{\arraystretch}{1.15}
\begin{tabular}{ll | r | r@{$\,\pm\,$}r @{\hspace{8pt}} | r@{$\,\pm\,$}r | c | r}
\hline\hline
Set & Method & $|\mathcal{U}|$ & \multicolumn{2}{c|}{$\langle n \rangle \pm \sigma$} & \multicolumn{2}{c|}{${\mathcal R_{0.97}} \pm \sigma$} & $\Delta\mathcal{U}$(\%) & $\Delta n$(\%) \\
\hline
\multirow{3}{*}{Set I}
& Random   & 400 K   & 87,549  & 106  & 96.39 & 0.07 & \multicolumn{1}{c|}{---} & \multicolumn{1}{c}{---} \\
& Halton-S & 290 K   & 86,792  & 103  & 96.46 & 0.06 & $27.5$ & \multicolumn{1}{l}{$[-1.10,\ -0.63]$} \\
& Halton   & 290 K   & 86,914  & 159  & 96.42 & 0.12 & $27.5$ & \multicolumn{1}{l}{$[-1.03,\ -0.42]$} \\
\hline
\multirow{3}{*}{\textcolor{black}{Set II}}
& \textcolor{black}{Random}   & \textcolor{black}{8000 K} & \textcolor{black}{482,806} & \textcolor{black}{214}  & \textcolor{black}{96.09} & \textcolor{black}{0.05} & \multicolumn{1}{c|}{\textcolor{black}{---}} & \multicolumn{1}{c}{\textcolor{black}{---}} \\
& \textcolor{black}{Sobol-S}  & \textcolor{black}{7045 K} & \textcolor{black}{478,235} & \textcolor{black}{343} & \textcolor{black}{96.05} & \textcolor{black}{0.07} & \textcolor{black}{$12$} & \multicolumn{1}{l}{\textcolor{black}{$[-1.06,\ -0.83]$}} \\
& \textcolor{black}{Sobol}    & \textcolor{black}{7045 K} & \textcolor{black}{477,660} & \textcolor{black}{166} & \textcolor{black}{96.06} & \textcolor{black}{0.04} & \textcolor{black}{$12$} & \multicolumn{1}{l}{\textcolor{black}{$[-1.14,\ -0.99]$}} \\
\hline\hline
\end{tabular}
\caption{Comparison of stochastic template banks generated using uniform random sampling and LDS: Halton and Sobol over the parameter spaces defined in Table~\ref{tab:sets}. \textcolor{black}{This table summarizes the Set~I and Set~II multi-realization comparisons; the Set~III recovery-curve comparison is presented separately in Fig.~\ref{fig:R097_vs_U}.} Banks are constructed using 10 independent realizations of a fixed $|\mathcal{U}|$ and tested against a common set of $100{,}000$ random injections per parameter space. Here, $\langle n \rangle \pm \sigma$ gives the mean and standard deviation of the number of templates, and $\mathcal{R}_{0.97} \pm \sigma$ denotes the mean recovery fraction above $\mathcal{FF}=0.97$. The quantity $\Delta \mathcal{U}$ represents the percentage reduction in proposal set size relative to random uniform sampling. $\Delta n$ gives the corresponding range in percentage reduction in template count, accounting for the $\pm\sigma$ uncertainties of both the LDS method and the random baseline. ``S'' denotes scrambled LDS.}
\label{tab:results}
\end{table}

In the 2D case, random uniform sampling achieves an average recovery fraction of $96.39 \pm 0.07\%$. Halton and scrambled Halton (Halton-S) sequences yield comparable or slightly higher recovery fractions of $96.42 \pm 0.12\%$ and $96.46 \pm 0.06\%$, respectively, while requiring $27.5\%$ fewer proposal points. 

\textcolor{black}{In the 3D case, random-uniform sampling achieves an average recovery
fraction of $96.09 \pm 0.05\%$, while Sobol and scrambled Sobol
(Sobol-S) achieve comparable recovery fractions of
$96.06 \pm 0.04\%$ and $96.05 \pm 0.07\%$, respectively, using
approximately $12\%$ fewer proposal points.}

\textcolor{black}{
The ten-realization results summarized in Table~\ref{tab:results} are
therefore consistent with the matched-recovery analysis above.
}

\begin{figure}[h]
    \centering
    \includegraphics[width=0.9\textwidth]{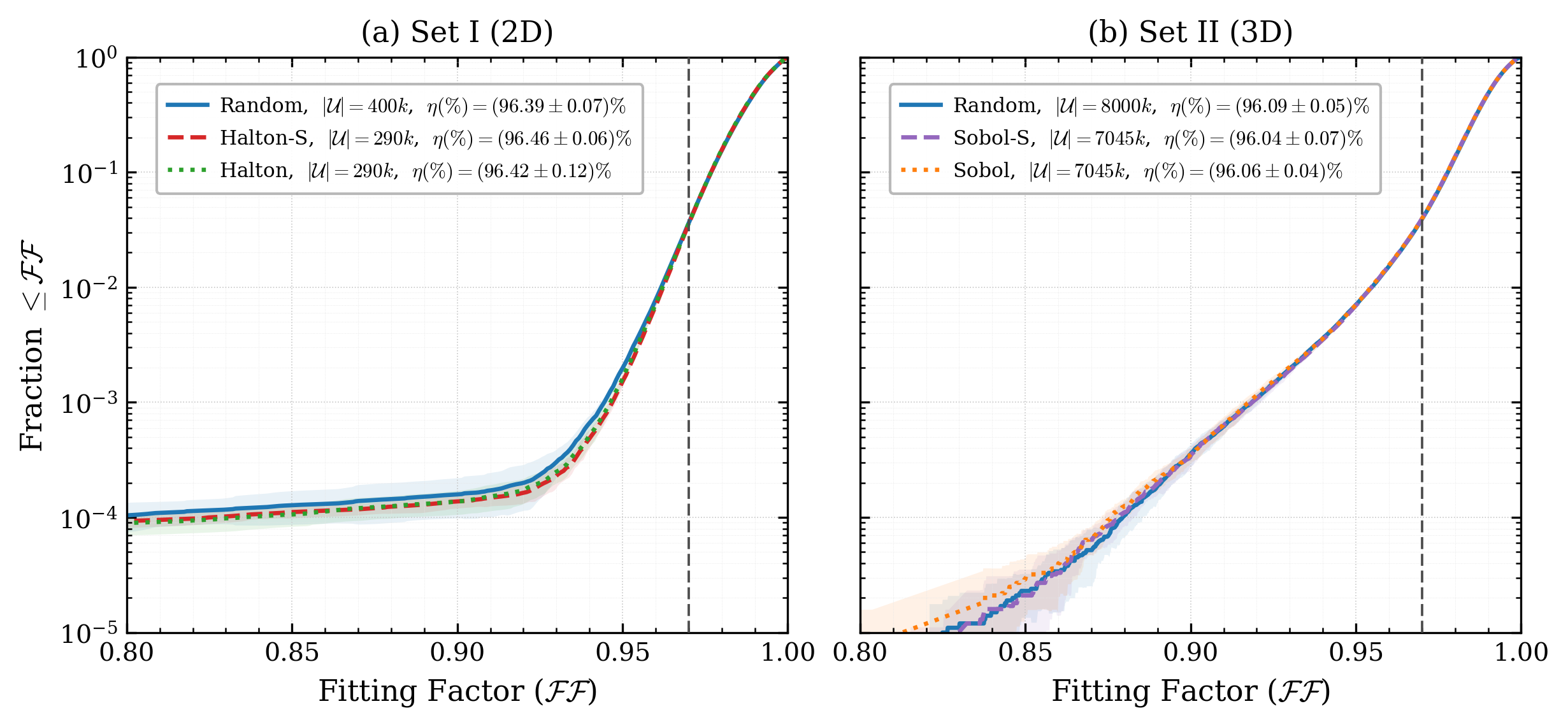}
    \caption{Cumulative distribution of the fitting factor $\mathcal{FF}$ for template banks constructed in Table~\ref{tab:results}, shown for Set~I/2D (left) and Set~II/3D (right). Each curve represents the mean CDF across 10 banks, constructed from 10 independent realizations of a fixed $|\mathcal{U}|$ and evaluated on a common set of $100{,}000$ random injections per parameter space. The shaded band shows the $\pm\sigma$ across realizations. The vertical dashed line marks the minimal match threshold $\mathcal{FF} = 0.97$, above which the recovery fraction ${\mathcal R_{0.97}}$ is computed.}
    \label{fig:cdf_plots}
\end{figure}

The corresponding cumulative distribution functions (CDFs) of the fitting
factors are shown in Fig.~\ref{fig:cdf_plots}.

\textcolor{black}{
The small fraction of Set~II injections that fall below the nominal minimal
match consists predominantly of marginal failures: about $82\%$ of these have
$0.95\leq\mathcal{FF}<0.97$. Only about $0.7\%$ of all injections have
$\mathcal{FF}<0.95$, while $\mathcal{FF}<0.90$ occurs for only about
$0.035\%$. The lowest-fitting-factor injections are preferentially
associated with parameter-space boundaries, particularly the equal-mass
boundary, the NS--BH transition at $3\,M_\odot$, and the upper
secondary-mass boundary $m_2=3\,M_\odot$.
}

We find that LDS sampling is a more efficient way to generate proposals for
the stochastic template-bank construction considered here. The final template
count changes only marginally, by about $1\%$ in both 2D and 3D, consistent
with the fact that the bank size is mainly set by the metric volume of the
target parameter space and the chosen minimal match.

The practical gain is instead in the size of the initial proposal set
$\mathcal{U}$. With fewer points to store and process, LDS sampling reduces
memory use, bookkeeping, and nearest-neighbour search overhead.

\textcolor{black}{To estimate the runtime impact, we compare the
bank-construction times at the selected values of $|\mathcal U|$ for
Sets~I and II.}

For Set I (2D), a reduction of $\sim 4.5\%$ in CPU time is observed, with bank generation taking 85.44 minutes for random uniform sampling, compared to 81.52 minutes for Halton and 81.50 minutes for scrambled Halton on a single CPU. 

\textcolor{black}{For Set II (3D), the banks were generated using a parallel implementation of the bank-generation code over 20 CPUs. The mean runtime decreases from $5.21\pm0.02$ hours for random-uniform sampling to $4.83\pm0.01$ hours for Sobol and $4.93\pm0.02$ hours for scrambled Sobol, corresponding to speed-ups of about $7.2\%$ and $5.3\%$, respectively.}

\textcolor{black}{The slab-wise pre-filter is not used for either of these 2D or 3D timing measurements; the observed speed-up therefore arises from the reduced proposal set size enabled by LDS sampling.} 
\textcolor{black}{The quoted runtime reductions refer to the final bank-construction runs at the selected values of $|\mathcal U|$ and do not include the separate recovery-curve calibration used to choose $|\mathcal U|$.}


These results show that the main advantage of LDS sampling lies in the
proposal stage: it reduces the number of initial points needed for comparable
coverage, while the final bank size remains largely set by the metric volume
and minimal match. The size of the numerical gain will depend on the
waveform model, PSD, and parameter-space boundaries, but the reason for the
improvement is generic to the method: better space filling than uniform
random sampling. \textcolor{black}{Accordingly, the numerical reductions in
$|\mathcal{U}|$ quoted here should not be interpreted as universal or
invariant under changes in the signal-manifold metric.}

\textcolor{black}{
The LDS and slab-filter improvements act at different stages of the
construction and are complementary: LDS sampling reduces the number of
proposals required to achieve a given recovery, while the slab-wise
pre-filter reduces the number of candidate templates passed to the exact
metric test during covering. The latter becomes particularly effective in
the 4D bank, where the stronger metric anisotropy makes the slab cut highly
selective.} Since the present study uses a computationally inexpensive
metric approximation for match evaluation, the benefit of this filtering
can be expected to increase when candidate comparisons require
computationally expensive waveform overlaps.


\section{DISCUSSION}
We generated and tested several template banks in \textcolor{black}{2D, 3D, and 4D} and found that
low-discrepancy sequence (LDS) sampling is more efficient for stochastic
template placement than random uniform sampling. 

The main bottleneck in these stochastic algorithms is the size of the
initial proposal set $\mathcal{U}$. Using the matched-recovery definition in
Eq.~\eqref{eq:lds_advantage}, LDS sampling reduces the required
$|\mathcal{U}|$ by approximately $27.5\%$ in 2D and $12\%$ in 3D.
\textcolor{black}{In the 4D eccentric equal-spin bank, Sobol sampling
achieves the same recovery as random-uniform sampling with about $17\%$
fewer proposals. The size of this reduction depends on the waveform model,
PSD, parameter-space boundaries, and signal-manifold geometry, and need not
vary monotonically with dimensionality. The 4D result therefore shows that
the LDS reduction in proposal-set size persists in a higher-dimensional
CBC bank.}


We also introduced a slab-wise coarse filtering strategy
(Sec.~\ref{sec:slabwise_filtering}) that accelerates candidate rejection
for highly anisotropic metrics in higher-dimensional applications.
\textcolor{black}{
To demonstrate both the LDS proposal reduction and the slab-wise prefilter
in a higher-dimensional setting, we use the 4D eccentric equal-spin toy bank
as a complete bank-construction test. Sobol sampling achieves the same recovery
as random-uniform sampling with about $17\%$ fewer proposals. Separately, the
single-slab pre-filter reduces the number of candidates passed to the exact
metric test by $99.53\%$, accelerates the covering stage by a factor of about
$2.9$, and gives a total bank-construction speed-up of
$1.27$--$1.28\times$, without changing the final bank. The slab-filter gain
is measured in the fixed-$|\mathcal U|$ ablation described above and is
therefore distinct from the matched-recovery reduction in proposal-set size.
}

\textcolor{black}{The top-down stochastic placement is inherently sequential,
since the acceptance or rejection of each proposal depends on the templates
already retained in the bank. Metric evaluations and spatial-query operations
can nevertheless be parallelized or partitioned, while independent
$|\mathcal U|$ trials and injection-based validation can be run concurrently.
A systematic study of parallel and distributed implementations is left to
future work.}


As expected, the final template count is only modestly reduced,
by about 1\% in both two and three dimensions since LDS
sampling does not change the underlying geometry of the placement problem.
At a fixed minimal match, the number of templates $N$ needed to cover the
$n$-dimensional intrinsic parameter space scales as
\begin{equation}
N \propto \frac{V_g}{(1-\mathcal{M}_{\min})^{n/2}},
\end{equation}
where $V_g = \int \sqrt{|g|}\, d^n\lambda$ is the metric volume of the
signal manifold and the proportionality constant depends on the placement
algorithm's coverage efficiency. Since this scaling is set by the manifold
geometry and the chosen minimal match, large reductions in bank size are not
expected from LDS sampling alone.

The advantage of LDS sampling lies instead in shrinking the oversampled
proposal set $\mathcal{U}$ needed for a given level of coverage. Because the
memory footprint and computational cost of stochastic and hybrid placement
algorithms are largely set by $\mathcal{U}$ and its associated search
structures (e.g.\ KD-trees), a smaller $|\mathcal{U}|$ translates directly
into lower memory use and faster bank construction. \textcolor{black}{The
fractional reduction in $|\mathcal{U}|$ need not increase with dimension.
However, as the proposal sets required for stochastic placement become larger,
even a comparable fractional reduction can yield a larger absolute reduction
in memory and proposal-processing cost.}

LDS sampling could also be incorporated into existing hybrid
placement methods, which combine  top-down stochastic placement with
lattice-based geometric placement in locally flat regions but still
typically begin from a uniformly sampled $\mathcal{U}$. Replacing this
initial draw with an LDS proposal set could reduce $|\mathcal{U}|$ at
comparable coverage, lowering memory and bookkeeping costs, with a possible
small gain in final template count. This would be especially useful in
optimized hybrid banks that rely on exact match calculations in parts of
parameter space~\citep{PhysRevD.99.024048} -- for instance high-mass
binaries, where the metric approximation can underestimate the true
mismatch neighbourhood, and a smaller proposal set means fewer expensive
match evaluations when marking proposals inactive. 
\textcolor{black}{Such reductions in proposal-set size may be especially useful
for future searches with higher-dimensional waveform families, such as
precessing and eccentric binaries~\citep{Gamboa2025, Phukon2025}, where the required
proposal sets can become very large.}

\appendix
\section{Metric Calculation}
\label{sec:appendix_metric}

In Section~\ref{sec:Parameter_space_metric}, we introduced the intrinsic
parameter-space metric $g_{ij}$ used for template-bank construction. We
summarize its calculation for the \IMRPhenomD{} waveform family.

The detector output may be regarded as a vector in a linear data space
equipped with the noise-weighted inner product
\begin{equation}
\langle a|b\rangle
=
4\,\mathrm{Re}
\int_{f_{\rm low}}^{f_{\rm high}}
\frac{\tilde a^{*}(f)\tilde b(f)}
{S_n(f)}
\,df,
\label{eq:inner_product}
\end{equation}
where frequency bins are weighted by the noise power spectral density
$S_n(f)$. For the metric calculations in this work, we take
$f_{\rm high}=f_{\rm ISCO}$, the Schwarzschild ISCO frequency, and use the
\aLIGOZeroDetHighPower{} noise power spectral density~\citep{LIGO-T0900288}.

\noindent
A family of normalized waveforms $\hat h(\lambda,\xi)$ forms a manifold
embedded in this data space, where $\lambda^i$ and $\xi^\mu$ are the
intrinsic and extrinsic parameters. The overlap between two nearby waveforms
is their match,
\begin{equation}
\mathcal M
=
\left\langle
\hat h(\Theta)
\middle|
\hat h(\Theta+\Delta\Theta)
\right\rangle,
\label{eq:A3}
\end{equation}
with $\Theta^A=(\lambda^i,\xi^\mu)$. Expanding about the maximum at
$\Delta\Theta^A=0$, where the first derivatives vanish, gives
\begin{equation}
1-\mathcal M
\simeq
-\frac{1}{2}
\left.
\frac{\partial^2\mathcal M}
{\partial\Delta\Theta^A
\partial\Delta\Theta^B}
\right|_{\Delta\Theta=0}
\Delta\Theta^A
\Delta\Theta^B.
\label{eq:A4}
\end{equation}

The Hessian of the match at $\Delta\Theta=0$ defines the Fisher matrix,
\begin{equation}
\Gamma_{AB}
=
-
\left.
\frac{\partial^2\mathcal M}
{\partial\Delta\Theta^A
\partial\Delta\Theta^B}
\right|_{\Delta\Theta=0},
\label{eq:A6}
\end{equation}
equivalently, the curvature of the log-likelihood at its maximum. 
Since
${\langle\hat h|\hat h\rangle=1}$ everywhere for normalized waveforms on the manifold, differentiating
this normalization twice gives $\langle\hat h|\partial_A\partial_B\hat
h\rangle=-\langle\partial_A\hat h|\partial_B\hat h\rangle$, so
Eq.~\eqref{eq:A6} becomes an inner product of waveform derivatives,
\begin{equation}
    \Gamma_{AB}=\langle\partial_A\hat h|\partial_B\hat h\rangle.
\end{equation}

\begin{figure}[t]
    \centering
    \includegraphics[width=0.975\textwidth]{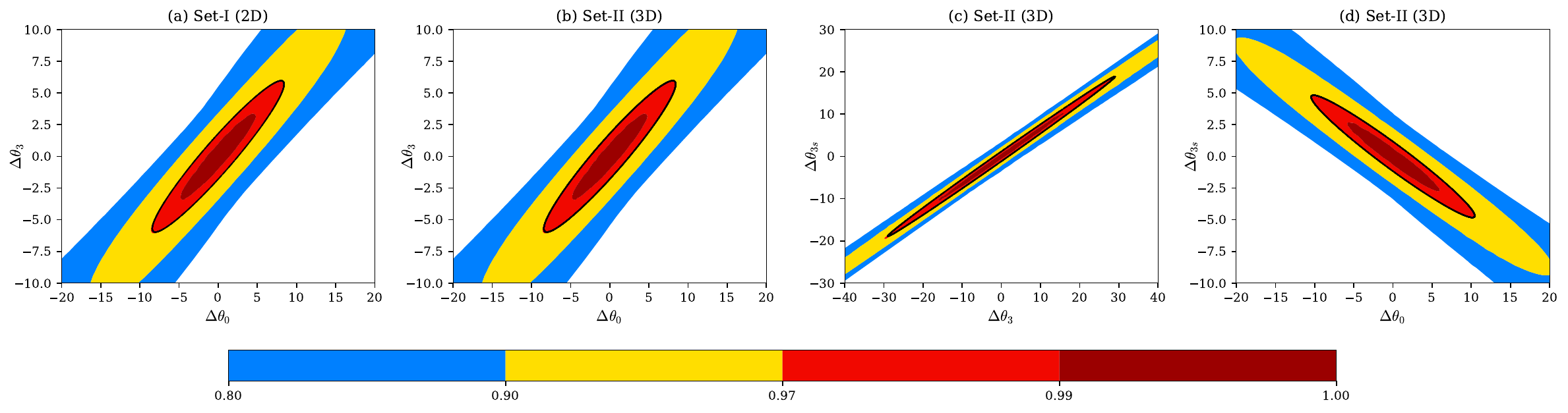}
    \caption{
This figure compares the metric approximation with the exact match as defined in Eq.~(\ref{eq:A3}). 
Panels (a) show contours of the numerically computed exact match in the $(\theta_0,\theta_3)$ plane at the central points of Set~I(Table~\ref{tab:sets}). 
Panels (b)--(d) show orthogonal slices of the exact-match contours in the three-dimensional parameter space $(\theta_0, \theta_3, \theta_{3s})$ at the center of Set~II (Table~\ref{tab:sets}). 
The corresponding metric ellipses are shown in black and are computed using Eq.~(\ref{eq:A9}). All black ellipses correspond to the contour at minimal-match $\mathcal{M_{\min}} = 0.97$.
}
    \label{fig:combined-metric-plot}
\end{figure}
Substituting Eq.~(\ref{eq:A6}) into Eq.~(\ref{eq:A4}),
\begin{equation}
1-\mathcal M
\simeq
\frac{1}{2}
\Gamma_{AB}
\Delta\Theta^A
\Delta\Theta^B.
\label{eq:A7}
\end{equation}

Geometrically, $\Gamma_{AB}$ is the pullback of the data-space metric,
Eq.~\eqref{eq:inner_product}, onto the manifold of normalized waveforms. For
frequency-domain waveforms of the form $\tilde h(f;\Theta) =
A(f;\Theta)\,e^{-i\Psi(f;\Theta)}$, such as
\IMRPhenomD{}~\citep{PhysRevD.93.044007}, this inner-product form reduces to
\begin{equation}
\Gamma_{AB}
\simeq
\frac{1}{\langle h|h\rangle}
\left[
\left\langle
\partial_A A
\middle|
\partial_B A
\right\rangle
+
\left\langle
A\,\partial_A\Psi
\middle|
A\,\partial_B\Psi
\right\rangle
\right].
\label{eq:A8}
\end{equation}

Templates are placed only on the intrinsic parameters $\lambda^i$, so the
extrinsic directions $\xi^\mu$ are projected out via
the Schur complement of the extrinsic block. Additionally, by using Eq.~\eqref{eq:A7} the
resulting intrinsic mismatch metric is
\begin{equation}
g_{ij}
=
\frac{1}{2}
\left(
\Gamma_{ij}
-
\Gamma_{i\mu}
\Gamma_{\mu\nu}^{-1}
\Gamma_{\nu j}
\right),
\label{eq:A9}
\end{equation}
where $i,j$ run over intrinsic and $\mu,\nu$ over extrinsic parameters. The
local mismatch between nearby templates then reduces to
$
1-\mathcal M
\simeq
g_{ij}
\Delta\lambda^i
\Delta\lambda^j,
$
and this is the metric, Eq.~\eqref{eq:A9}, used for template-bank placement.


In Fig.~\ref{fig:combined-metric-plot}, we assess the validity of this approximation by comparing the resulting metric ellipses with contours from exact match calculations at representative points in the 2D and 3D parameter spaces.

\vspace{.5em}
{\centering
    \raisebox{0.3ex}{\rule{0.15\textwidth}{0.7pt}}\;\;
    \(\blacksquare\)\;\;
    \raisebox{0.3ex}{\rule{0.15\textwidth}{0.7pt}}
\par}

\section*{Data availability statement}

The numerical data supporting the results presented in this work are
publicly available on Figshare at
\url{https://doi.org/10.6084/m9.figshare.33869146}.

\ack{
\textcolor{black}{The authors thank the two anonymous referees for their careful reading and thoughtful comments, which helped considerably improve the clarity and scope of the manuscript.} The authors thank Sudhir Gholap for carefully reading the manuscript and giving helpful suggestions to improve the presentation and clarity. We also thank Abhishek Sharma and Divya Tahelyani for useful discussions and comments, and acknowledge support from IIT Gandhinagar, where the computations reported in this work were performed on the \texttt{noether} workstation of the Physics Department.
Tarun Kumar would also like to thank Aayushi Jain, Akash Sahoo, and Ankit Mishra for their help during the initial stages of preparing the manuscript. Finally, the authors thank IIT Gandhinagar's computer support staff for their prompt assistance in resolving computing-related issues.

}

\bibliography{references}
\bibliographystyle{unsrtnat}

\end{document}